\documentclass{article}%
\usepackage{amsmath}%
\usepackage{amsfonts}%
\usepackage{amssymb}%
\usepackage{graphicx}
\usepackage{xcolor}

\begin{document}
	
	\vspace{32pt}
	\begin{center}
		{\textbf{\Large NEUTRINOS AND STRONG INTERACTIONS: QCD (and BEYOND?)\footnote{Invited talk at LISHEP 2021, Rio de Janeiro, July 2021}}}
		\vspace{40pt}
		
		C.A.~Garc\'\i a Canal\\
		\vspace{12pt}
		\textit{IFLP/CONICET and Departamento de F{\'\i}sica}\\ \textit{Universidad Nacional de La Plata, C.C.67, 1900, La Plata, Argentina}
		
	\end{center}
	
	\vspace{40pt}
	
	\date{\today}%
	
	\begin{abstract}
		
		In this talk the answer to the question: what neutrinos have to do with QCD, the present theory of strong interactions? is presented.
		The answer is positive as some strong arguments on this assertion that are presented bring out.
		
	\end{abstract}

	\section{Introduction}

	The first question to pose and to answer seems to be: what has to do a neutrino that only participates of the weak interaction with QCD the theory
	of strong interactions?  Through this talk we hope that it will became clear that, no doubt, the neutrino has a lot to do with QCD.
	
	The history (or perhaps I should write story...) starts in the seventies when the quark structure of nucleons, the proton in particular,
	was confirmed by the experimental data obtained in the Gargamelle bubble chamber at CERN acted by neutrino beams and it continues nowadays, mainly in connection with the IceCube detected events initiated by very energetic cosmic neutrinos in this case.
	
	The experimental study of electron inelastic scattering at SLAC \cite{SLAC} and the results obtained at CERN \cite{CERN1} in the bubble chamber Gargamelle, in the seventies, detecting neutrino reactions, gave strong support to the image of the nucleon containing particular constituents, called partons, that afterwards were identified with the quarks of the $SU(3)_{flavor}$ spectroscopy hypothesis. Initially, data confirm the scaling behavior of deep inelastic cross sections, and moreover, that the electric charge assigned to quarks was confirmed.
	
	Gargamelle apparatus owes its name to the mother of Gargantua, the character of Rabelais' series of bookas: "Gargantua et Pantagruel", mainly due to the big size of the bubble chamber. It measured $4.8 \, m$ long and $2\, m$ in diameter, weighed $1000\, T$ and was filler with $12 m^3$ of heavy-liquid freon ($CF_3Br$). In July 1973, the Gargamelle collaboration announced the detection of weak neutral currents ~\cite{Garga1} being the first experimental indication of the existence of the $Z_0$ weak boson, and for that reason a verification of the electroweak theory based upon the
	$SU(2) \otimes U(1)$ gauge symmetry. Unfortunately, in 1979 an irreparable crack appeared in the bubble chamber, and Gargamelle was decommissioned. It is at present part of the "Microcosm" exhibition at CERN. (At that time, the cracking was particularly impressive because the tank of Gargamelle was builded by the same industry that constructed nuclear reactor tanks...). This cracking is directly connected with my participation in the physics of Gargamelle. Together with the Brazilian colleagues Mar\'{\i}a Beatriz Gay and Jose Martins Sim$\tilde{o}$es we were contacted in Strasbourg by one of the experimental collaborations working in Gargamelle, the Aachen-Bergen-Brussels-London-Strasbourg one, that wanted us to compute the expected number of same sign dimuon events in neutrino-nucleon scattering to be expected taking into account the small time of recording of data due to the decommission of the bubble chamber. By using the QCD improved parton model we concluded that the collaboration could have registered at most $5$ dimuon events~\cite{ABBLS}. If they were democratic, each member of the collaboration should get one event...
	
	In 1961 Gell-Mann, based on the symmetry group $SU(3)_{flavor}$, introduced the Eightfold Way scheme, which classified the hadrons into representations of that group. Mathematically, Gell-Mann's $SU(3)$ model has a fundamental representation of dimension $3$ that perhaps would imply the existence of three more elementary particles, that were called "quarks". At the same time, Zweig in a paper of 1964,  proposed that mesons and baryons are formed from a set of three fundamental particles that he called aces.Both Gell-Mann's quarks and Zweig's aces had exotic electrical charges equal to $-1/3$ and $+2/3$ of the electron charge. The situation, at that time (1966) was such that Gell-Mann himself stated: "..the idea that mesons and baryons are made primarily of quarks is hard to believe..."
	
	As it was previously mentioned, in 1968,  electron-proton scattering experiments by the MIT-SLAC collaboration at SLAC revealed the inner structure of protons. The experiment was based on electrons colliding protons and the detection of the electrons that emerge. The scattering results allows one to conclude that the interaction were caused by point-like, incoherent constituents inside the protons. In the subsequent years, these results were confirmed and combined with those from neutrino-scattering in the Gargamelle bubble chamber at CERN, showing conclusively that these point constituents have charges of $+2/3\,e$ and $-1/3\,e$.
	
	A brief reminder of the background for this result ~\cite{GCGS}:
	In the case of electromagnetic deep inelastic scattering (DIS) one has for the structure function in the quark-parton model
	\[
	F_2(x) = \sum_i\,e_i^2\,q_i(x)
	\]
	where $e_i$ stands for the electric charge of the $i$ quark involved with quark distribution function $q_i$.
	Consequently, for a proton target
	\[
	F_2^{\gamma p}= \frac{4}{9}\,(u + \bar{u}) + \frac{1}{9}\, (d + \bar{d}) + \frac{1}{9}\, (s + \bar{s})
	\]
	and a similar one for $F_2^{\gamma n}$ with the replacement $u\leftrightarrow d$ in the case of a neutron.
	
	On the other hand, when the DIS is initiated by neutrinos, the interactions present: $\nu\,d\rightarrow\ell^-\,u$ and
	$\nu\,\bar{u}\rightarrow\ell^-\,\bar{d}$ give rise to
	\[
	F_2^{\nu p} = 2  (d + \bar{u})
	\]
	and again the replacement $u\leftrightarrow d$ allows one to write $F_2^{\nu n}$
	For the case of an isoscalar target $N$, the previous expressions ends in the relation
	\[
	\frac{F_2^{\nu N}}{F_2^{\gamma N}} = \frac{\frac{5}{9}\,(u + \bar{u}+d + \bar{d}) + \frac{2}{9}\,(s + \bar{s})}
	{2\,(u + \bar{u}+d + \bar{d})} \geq \frac{5}{18}
	\]
	
\begin{figure}[h]
		\begin{center}
			\includegraphics[scale=1]{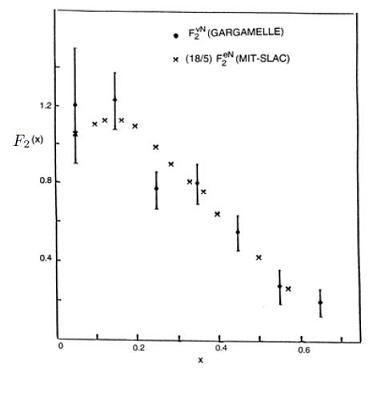}
			\caption{Comparison of SLAC-Gargamelle DIS data}
			\label{518}
		\end{center}
	\end{figure}

	This number $5/18$ that comes from the electric charges of quarks, was experimentally obtained comparing Gargamelle data with
	the SLAC data, as it is shown in Fig.\ref{518}

	Consequently, neutrinos gave physical sense to the fractional charge of quarks, now identified with the nucleon components This also allows stating that the electron and neutrino scattering measurements found the same structure. These detected components are the basic degrees of freedom for building up a quantum field theory of strong interactions based upon the gauge symmetry $SU(3)_c$, QCD.

	Much water has gone under the bridge since Gargamelle's successes and we are not trying here to summarize the recent history of the neutrino physics~\cite{fundamental}.  We are only interested in presenting one of the most recent approaches of using very high energy neutrino data to eventually confirm at these energies QCD or detect new physics beyond the Standard Model. It is important to realize that even if neutrinos participate directly only of the weak interactions (leaving aside gravity of course) they are very useful as probes of the structure of hadrons and in particular of proton structure, as it was remarked above.
	
	\section{Neutrino-Nucleon Interaction}
	
	In order to prepare the presentation of the impact of very high energy neutrinos detected by IceCube, we summarize here the most relevant equations
	connected with the neutrino-nucleon interaction, mainly in the deep inelastic (DIS) regime.
	
	The kinematics of neutrino-nucleon scattering is described in terms of the momentum transferred measured by $Q^2$, the Bjorken $x=Q^2/2\,m_N\,\nu$, and the inelasticity $y = Q^2/sx = \nu/E$
	that measures the energy transfer $\nu = E_{\nu, f}-E_{\nu, i}$ between the neutrino and the nucleon
	and with $s$  being the square of the center-of-mass
	energy.  The cross-section for CC neutrino (and antineutrino) scattering
	on isoscalar nucleon targets is given
	by~\cite{Devenish:2004pb}
	\begin{equation}
		\sigma_{{\rm CC},0} =\int_0^1 dx \int_0^{xs} dQ^2
		{d^2 \sigma^{\nu (\bar \nu) N} \over dx \ dQ^2}\, ,
		\label{sigma1}
	\end{equation}
	where
	\begin{eqnarray}
		{d^2\sigma^{\nu (\bar \nu) N} \over dx \ dQ^2} & = &
		{G_\mathrm{F}^2 \over 2\pi x}
		\bigg({m^2_W \over Q^2 + m^2_W}\bigg)^2
		\bigg[Y_+\, F_2^{\nu (\bar \nu)} (x, Q^2) \nonumber \\
		&  &-  y\, F_{\rm L}^{\nu (\bar \nu)}  (x, Q^2) + Y_-\,
		xF_3^{\nu(\nu)} (x, Q^2)\bigg]
		\label{sigma2}
	\end{eqnarray}
	is the differential cross-section given in terms of the structure
	functions $F_2^{\nu (\bar \nu)},$ $F_{\rm L}^{\nu (\bar \nu)}$ and
	$xF_3^{\nu (\bar \nu)}$, and $Y_+ = 1 + (1-y)^2$, $Y_- = 1 -
	(1-y)^2$. Here, $G_{\rm F}$ is the Fermi constant and $m_W$ is the
	$W$-boson mass. At leading order in perturbative QCD (LO), the
	structure functions can be given in terms of parton distributions
	\[
	F_2^{\nu (\bar \nu)}
	= x \big[\sum_i \alpha_i q_i(x,Q^2) + \sum_j \alpha_j \bar{q}_j(x, Q^2)\big]
	\]
	\[
	xF_3^{\nu (\bar \nu)} =
	x \big[ \sum_i  \beta_i q_i(x, Q^2) + \sum_j \beta_j \bar{q}_j(x, Q^2)\big]
	\]
	and, at LO
	\[
	F_{\rm L}^{\nu (\bar \nu)} =0
	\]
	For neutrinos, $i = u,d,s,b$ and $j= u,d,c$,
	with $\alpha_i = \alpha_j = \beta_i  = 1$ for $u,d$; $\alpha_i=\alpha_j = \beta_i = 2$ for
	$s,b$; $\beta_j = -1$ for $u,d$; $\beta_j = -2$ for $c$ quarks.  For antineutrinos,
	$i = u,d,c$ and $j = u, d,s,b$, with $\alpha_i = \alpha_j = \beta_i  = 1$ for $u,d$; $\alpha_i=\alpha_j = \beta_i = 2$ for
	$c$; $\beta_j = -1$ for $u,d$; $\beta_j = -2$ for $s,b$ quarks.
	
	The NC cross sections on isoscalar targets via $Z$ are given by expressions
	similar to (\ref{sigma1}) and (\ref{sigma2}), with the $W$ propagator
	replaced by the $Z$ one. For NC interactions the LO
	expressions for the structure functions are given by
	\[
	F_2^{\nu (\bar
		\nu)} = x \big[ \sum_i \alpha_i [q_i(x,Q^2) + \bar{q}_i(x, Q^2)] +
	\sum_j \alpha_j [q_j(x,Q^2) + \bar{q}_j(x, Q^2) ]
	\]
	\[
	+  \sum_k \alpha_k[
	q_k(x,Q^2) + \bar{q}_k(x, Q^2)] \big]
	\]
	and
	\[
	xF_3^{\nu (\bar \nu)} =
	\sum_i x (v_ua_u + v_d a_d) [ q_i(x, Q^2) - \bar{q}_i(x, Q^2)]
	\]
	where $i = u,d$, $j= s,b$, $k = c$, $\alpha_i = (a_u^2 +v_u^2 +a_d^2
	+v_d^2)/2$, $\alpha_j = a_d^2 +v_d^2$, and $\alpha_k =a_u^2 +v_u^2$,
	with $v_u$, $v_d$, $a_u$, $a_d$ the NC vector and axial couplings of
	$u-$ and $d$-type quarks.
	
	\section{Experimental Information}
	
	High-energy neutrinos are unique messengers of information of very far-away physics phenomena.
	For this reason, they could be eventually considered as an informant of physics beyond the Standard Model.
	The event rates that one could registered are a combination of neutrino flux and the corresponding cross section.
	In order to avoid the astrophysical uncertainties implied, the concomitant consideration of up-going and down-going
	events is in order. By using this combination one can disentangle physics from astrophysics \cite{Luis}.
	This is precisely the job that IceCube collaboration ~\cite{IceCube}, the neutrino telescope installed in
	the Amundsen-Scott station of the South Pole, is doing for us. This is precisely a counting experiment.
	
	We have investigated the sensitivity of the
	present and future South Pole neutrino-detection experiments,
	in particular on the neutrino-nucleon cross-section \cite{nos}. This was triggered by the detection in IceCube of sensible
	neutrino event rates at PeV energies ~\cite{ICe,Sta}. In fact, the IceCube Neutrino Observatory at the South Pole,
	which detects Cherenkov light from charged particles produced in neutrino interactions,
	firmly established the existence of an astrophysical high-energy neutrino component. They particularly refer to the so-called
	$HESE$-events (High Energy Starting Events), those that start inside the detector and in this way the $\mu$-background is
	substantially reduced. For energies above $60\,TeV$ the measured flux is
	\[
	\phi_0(E_\nu)=  2.2 \times 10^{-18}~\left(\frac{E_\nu}{100~{\rm
			TeV}}\right)^{-2.58}~(\mathrm{GeV\,s\,sr\,cm^{2}})^{-1}
	\]
	
\begin{figure}[h]
		\begin{center}
			\includegraphics[scale=1]{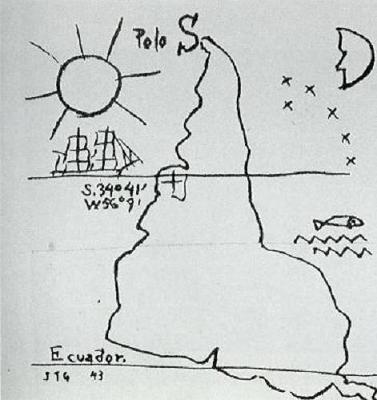}
		\end{center}
		\caption{Our South pictured by Joaqu\'{\i}n Torres Garc\'{\i}a}
		\label{STG}
	\end{figure}

\begin{figure}[h]
		\begin{center}
			\includegraphics[scale=0.7]{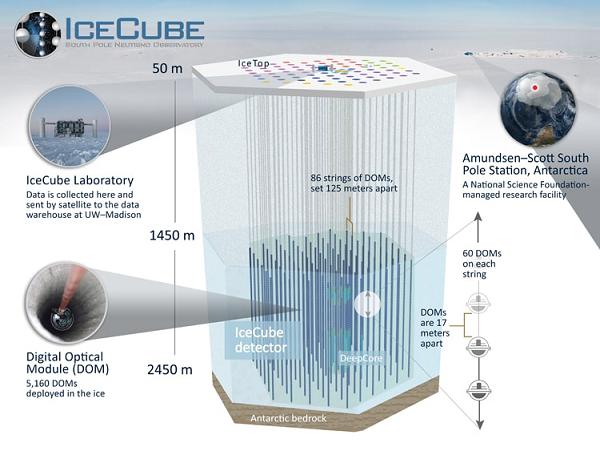}
			\caption{Sketch of the IceCube facility}
		\end{center}
		\label{IC1}
	\end{figure}

	Now is mandatory to go "up to the South" (see Fig.\ref{STG}) to learn about IceCube as presented in Fig.\ref{IC1}
	
	The IceCube Collaboration recent
	measurement of the neutrino-nucleon cross section~\cite{Aartsen:2017kpd} in
	the energy region $6.3 < E_\nu/{\rm TeV} < 980$ is
	\[
	\sigma_{\nu N} = \sigma_{\rm SM}  \times \left[1.30^{+0.21}_{-0.19} {\rm (stat.)} ^{+0.39}_{-0.43}
	{\rm (syst.)} \right] \,,
	\]
	where $\sigma_{\rm SM}$ stands for the
	Standard Model prediction~\cite{CooperSarkar:2011pa}.
	The energy
	dependence of the cross section was also determined ~\cite{Bustamante:2017xuy}. It has already been proposed the upgrade to
	the IceCube-Gen2~\cite{Aartsen:2014njl} to work technologically
	at the level of IceCube. A mere scaling of the corresponding dimensions allows one to estimate that IceCube-Gen2
	will have an order of magnitude larger aperture than IceCube, providing an expected improvement in the precision of measurements of
	the neutrino-nucleon cross-section to be comparable to the perturbative QCD related to
	collider data. Consequently, our main goal was to quantitatively infer the sensitivity of IceCube, present and future, for $\sigma_{\nu N}$ in an energy region well beyond that available for neutrino beams.
	
	The three neutrino flavors produces distinctive signal when they interact in
	ice producing the Cherenkov light detected by
	the IceCube.  The charged current
	interaction of $\nu_e$ triggers an electromagnetic cascade  (or shower) $S$
	producing a quite spherically
	symmetric signal, and therefore exhibits a low angular resolution but a relatively precise
	measurement of the $\nu_e$ energy. The
	situation is different for CC interaction $\nu_{\mu}$ induced
	events. In this case, the secondary muons travel relatively unhurt
	through the ice leaving tracks $T$ that point nearly in
	the direction of the original $\nu_{\mu}$, allowing a high angular resolution but the energy
	deposited represents only a lower bound of the $\nu_{\mu}$ energy. Certainly, misclassification is possible.
	The topology of events produced by
	different neutrino flavors and interactions, are indicated in
	Table~\ref{tab:topology}.
	
	\begin{table}[h]
		\caption{Event topology for each neutrino flavor.}
		\begin{tabular}{c|ccc}
			\hline
			\hline
			~~~Interaction type~~~& ~~~~~~$e$~~~~~~ & ~~~~~~$\mu$~~~~~~ & ~~~~~~$\tau$~~~~~~ \\\hline CC &$S$&$
			T$&$S$\\ NC &$S$&$S$&$S$\\ \hline \hline\end{tabular}\label{tab:topology}
	\end{table}
	
	The rates at IceCube for down- and up-going events have been
	found~\cite{Marfatia:2015hva} to scale respectively as $\Gamma_{\rm down} \propto
	\phi \ \sigma_i$ and $\Gamma_{\rm up} \propto \phi \ \sigma_i/\sigma_a$, where
	$\phi$ is the neutrino flux, $\sigma_i$ is the cross section for the
	interaction that produces the event ($i \in \{{\rm CC}, {\rm NC}\}$), and $\sigma_a$ is the attenuation
	cross section, including all the effects due to the fact that neutrinos go through the Earth.
	
	For a given flux $\phi$ and cross sections $\sigma_{i}$ and
	$\sigma_a$, the expected number of up-going events of a flavor $\alpha$
	produced by a charged or neutral current interaction may be expressed as
	\begin{subequations}
		\begin{equation}N_u^{i,\alpha}\equiv \tilde
			N_u^{i,\alpha}\frac{\phi}{\phi_0}\frac{\sigma_i/\sigma_a^\alpha}{\sigma_{i,0}/\sigma_{a,0}^\alpha},\end{equation}\label{eq:1}
		and for down going events,
		\begin{equation}
			N_d^{i,\alpha}\equiv \tilde
			N_d^{i,\alpha}\frac{\phi}{\phi_0}\frac{\sigma_i}{\sigma_{i,0}},\label{eq:rates0}\end{equation}
	\end{subequations}
	with  $i\in\{{\rm
		CC},{\rm NC}\}$ and where the $\tilde N$-constants are obtained assuming that the flux and
	cross sections are equal to the reference values, $\sigma_{i,0}$ and $\sigma_{a,0}$.
	
	\section{Likelihood Analysis}
	
	The tool for the analysis we have proposed is the Likelihood estimation.
	Likelihood states how probable a set of observations are given particular values of
	statistical parameters.
	To go ahead one introduces the set $\theta\doteq \{F,H\}$ of parameters involved in the data analysis
	defined by
	\begin{equation}
		\phi = F\,\phi_0\,\,\,\,;\,\,\,\,\sigma_{tot} = H\, \sigma_{tot;0}
	\end{equation}
	that are relevant to vary $\sigma_{CC;0}$ and $\sigma_{NC;0}$.
	
\begin{figure}[h]
		\begin{center}
			\includegraphics[scale=1]{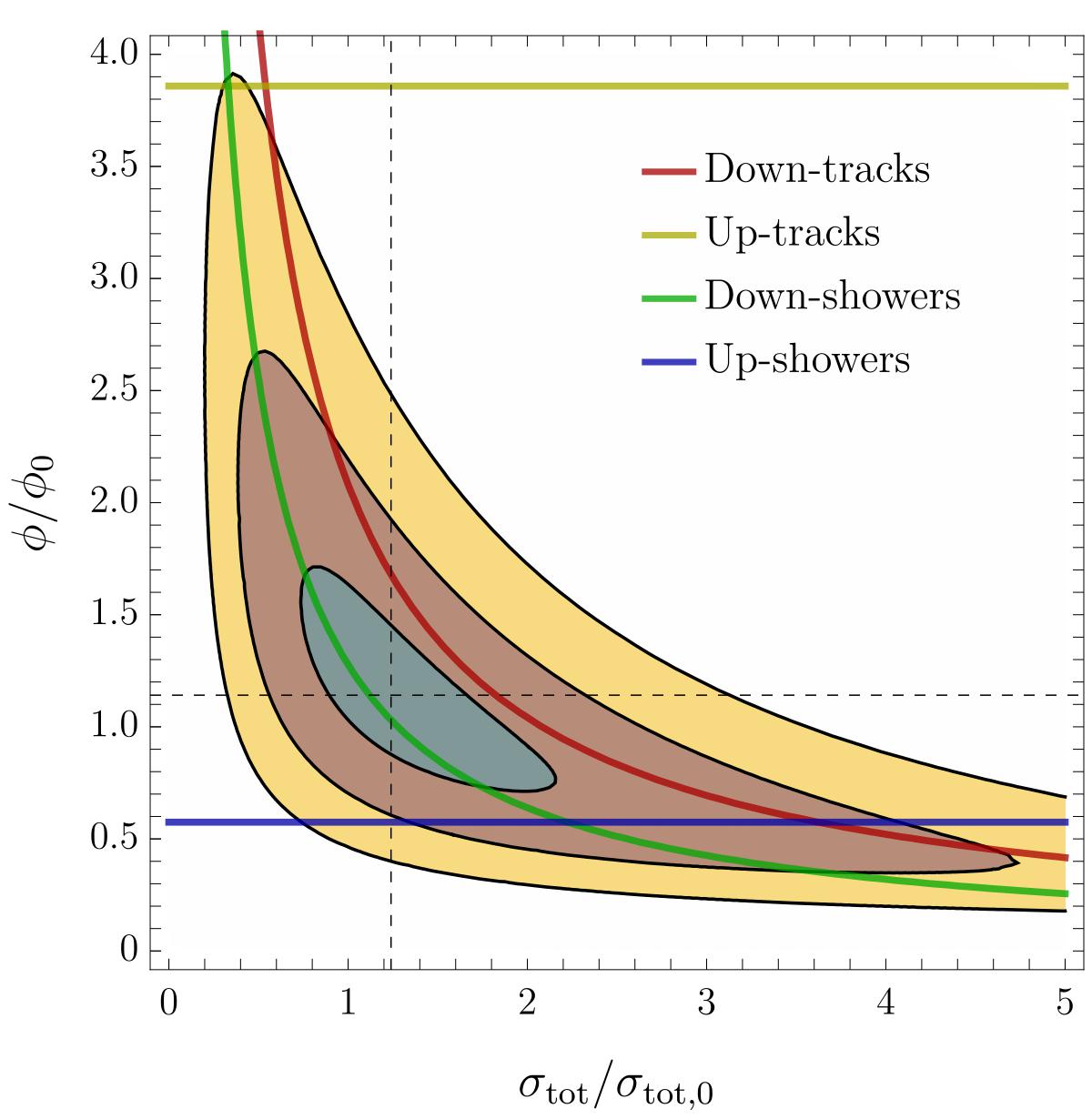}
		\end{center}
		\caption{1, 3, and $5\sigma$ confidence contours for $(F,H)$ for
			scaled total cross section $\sigma_{\rm tot}$ and flux $\phi$ with
			respect to their reference values $\sigma_{{\rm tot},0}$, $\phi_0$.}\label{fig:scale}
	\end{figure}

	Now, maximizing $L(\theta)$, the likelihood, allows one to estimate the most likely values of the parameters $F$ and $H$.
	This is obtained for the pair of values
	\begin{equation}\left\{\begin{array}{rcl}
			H&=&1.16^{+0.51}_{-0.34}\, (1\sigma\,\mbox{C.L.}),\\[.2cm]
			F&=&0.86^{+0.27}_{-0.22}\,
			(1\sigma\,\mbox{C.L.}).\end{array}\right.\label{eq:scaledr}\end{equation}
	In Fig.~\ref{fig:scale} we show the confidence contours and the associated curves in the $F-H$ plane for each
	event type that would produce the observed number of events of each
	type.
	
	Therefore, the cross-section is consistent at $1\sigma$ with the value from QCD guided by HERA data with some statistics
	limited by an order of $37\%$.
	
	The next step in our analysis includes $\sigma_{CC}/\sigma_{NC}$ as a free parameter. This is guided by the fact that
	in general, all new physics processes increases the neutral current cross section via, for example, gravitons,
	strings, sphalerons, etc. \cite{cornet}, i.e, as non-perturbative SM effects.
	
	We have used the
	complete set of ($ S\,+ \,T$) HESE data to find a bound to the
	rise of $\sigma_{\rm NC}$. Because the low number of
	data and the large
	uncertainties in arrival direction of shower events we have
	integrated over the angular distribution. Note that the analysis
	presented herein is complementary to those
	reported in~\cite{Aartsen:2017kpd,Bustamante:2017xuy} as it test a
	different region of the neutrino-nucleon cross section parameter
	space. Indeed, the likelihood fit given in (\ref{eq:scaledr12}) provides
	the first clear constraint coming from IceCube data on beyond the Standard Model
	phenomena.
	
	We have done again our analysis but keeping the ratio
	$\sigma_{\rm CC}/\sigma_{\rm NC}$ as a free parameter in the
	likelihood function instead to mantain it fixed to the value
	$\sigma_{\rm CC}/\sigma_{\rm NC} = 3$ that is expected in the Standard Model.
	
	We write the total neutrino-nucleon cross section as
	$\sigma_{\rm tot}=\sigma_{\rm CC,0}+\sigma_{\rm NC}$. Instead of
	considering the the parameter $H$ as the one of interest,
	we perform the analysis to constrain the ratio $H_{\rm NC}\equiv \sigma_{\rm NC}/\sigma_{\rm NC,0}$. Following a process similar to that
	previously used, maximizing the likelihood for the parameters $\boldsymbol\theta=\{F,S_{\rm NC}\}$  provides the values
	
	\begin{equation}\left\{\begin{array}{rcl}
			S_{\rm NC}&=& 0.00^{+0.27}_{-0.00}\, (1\sigma\,\mbox{C.L.}),\\[.2cm]
			F&=&1.16^{+0.20}_{-0.18}\,
			(1\sigma\,\mbox{C.L.}).\end{array}\right.
		\label{eq:scaledr12}
	\end{equation}
	
	\begin{figure}[h]
		\begin{center}
			\includegraphics[scale=1]{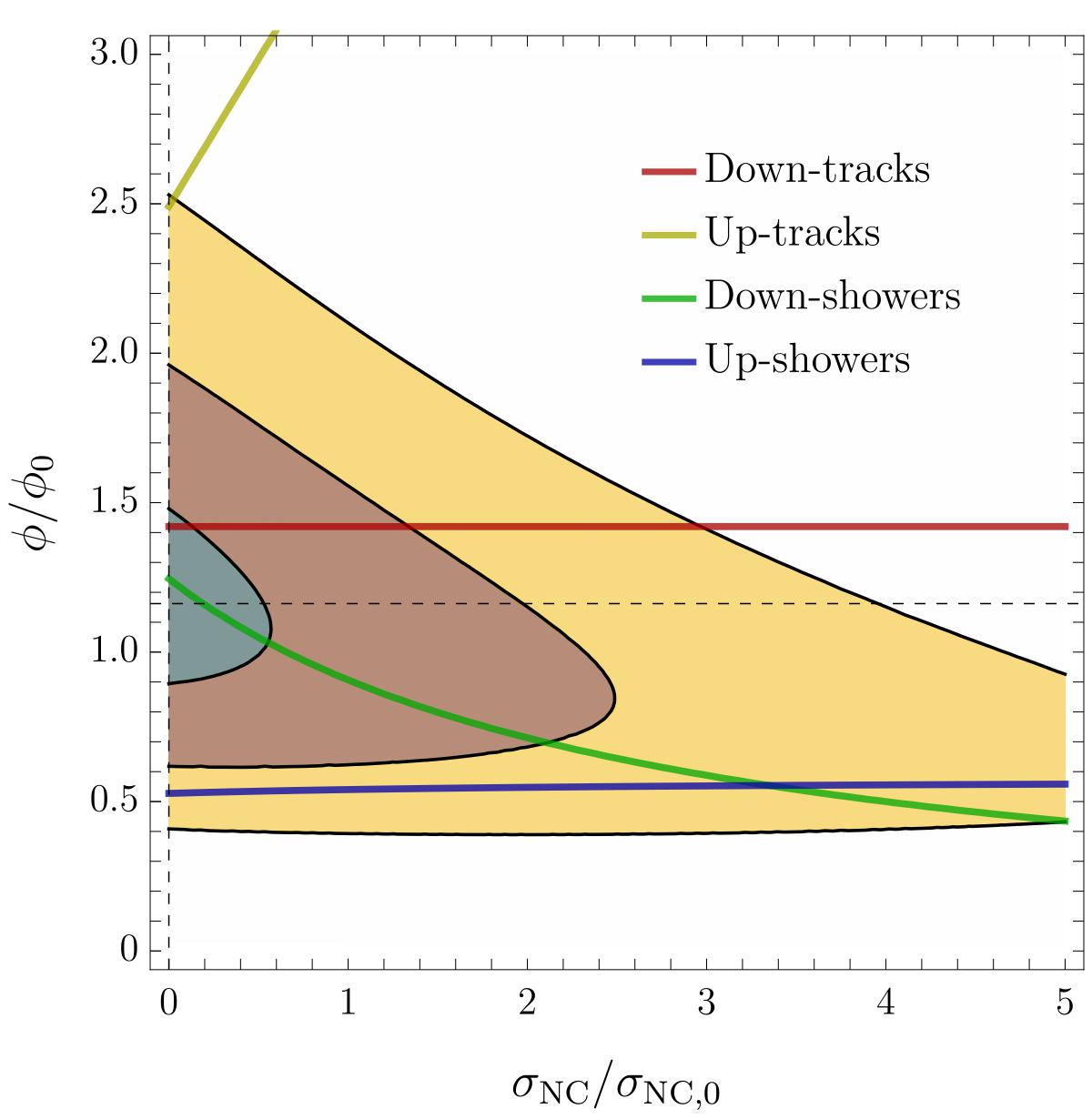}
		\end{center}
		\caption{1, 3, and $5\sigma$ confidence contours in the $(F,H_{\rm SN})$  plane.
			\label{fig:scale_nc}}\end{figure}
	
	In Fig.~\ref{fig:scale_nc}  the confidence contours and the
	associated curves in the $F - S_{\rm NC}$ plane for each event type
	that would produce the observed number of events of each type is presented.
	
	We can conclude that $H_{\rm NC}>1$ is excluded at $2\sigma$
	level. Consequently, there is a consistency with the value of QCD coming from HERA data.
	This result also is a constraint from IceCube data on non-perturbative SM phenomena, such as sphaleron transitions,
	that remain almost unconstrained by LHC data~\cite{Sirunyan:2018xwt}.

	\section{Looking Ahead}
	
	Studies for the upgrade to IceCube-Gen2 high-energy array are going on
	~\cite{Aartsen:2014njl}. The planned instrumented volume is around
	$10~{\rm km}^3$ and will lead to larger neutrino
	detection rates, in all neutrino flavor and detection channels. A
	rapid estimate indicates about an order of magnitude increase per year in
	the exposure. The sample size of events can be obtained by simply
	scaling the instrumented volume.

\begin{figure}[h]
		\begin{center}
			\includegraphics[scale=1]{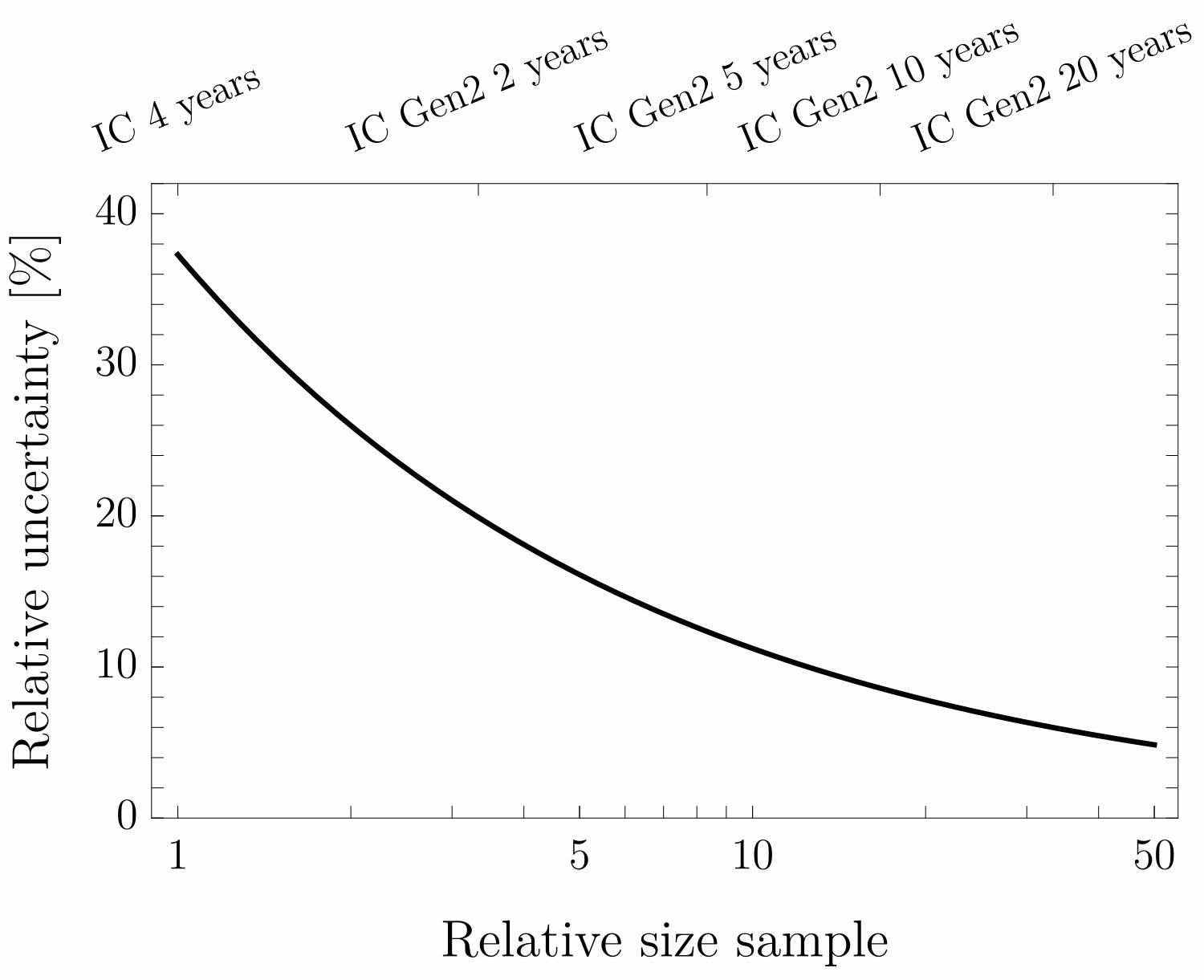}
		\end{center}
		\caption{Evolution of the cross section precision measurement.}\label{fig:accuracy}
	\end{figure}

	To get information on the sensitivity of IceCube-Gen2 to probe strong dynamics,
	we generate random samples of events
	following the Poisson distribution, with the parameters for a
	scaled total cross section found in the IceCube data analysis. In 10 years of observation
	IceCube-Gen2 will collect about 500 neutrino events in the energy
	range $0.1 \lesssim E_\nu/{\rm PeV} \lesssim 2$, and about 1000 events in 20
	years. Thus we adopt 20 and 40 as the representative multiplicative
	factors associated with these data samples. Using the high-energy and
	high-statistics sample to be collected by IceCube-Gen2, we perform the
	same likelihood analysis as with the real data. The precision on the cross section determination
	would be 7.9\% and 5.5\% for $\sim500$ and $\sim1000$ events, respectively. This
	precision is comparable to that obtained in perturbative QCD
	calculations guided by HERA data.  Detailed evolution of the
	uncertainty with sample sizes is illustrated in
	Fig.~\ref{fig:accuracy}.
	
	We can also dream an IceCube-like detector of $100~{\rm
		km}^3$, specifically designed to probe strong dynamics. In this case the $1\sigma$ contour regions
	could reach a precision of less than 2\% level.
	
\begin{figure}[h]
		\begin{center}
			\includegraphics[scale=0.7]{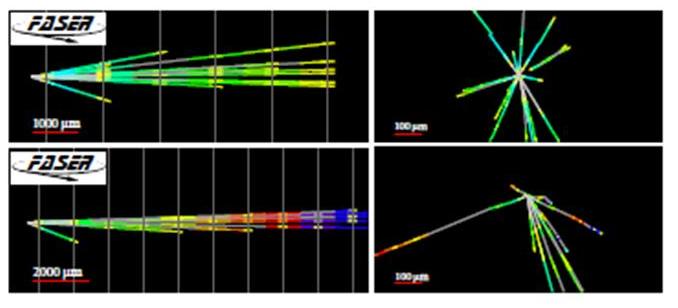}
		\end{center}
		\caption{Event displays of two of the neutral vertices in the y-z projection longitudinal to
			the beam direction (left) and in the view transverse to the beam direction (right), taken from Ref.\cite{FASER}}\label{fig:FASER}
	\end{figure}

	Before ending, we refer to the FASER$\nu$ collaboration at the CERN Large Hadron Collider. This is an experiment designed to  detect collider neutrinos directly, in order to analyze the neutrino cross-section at the TeV energies of LHC, a region of energies not yet studied. On May 13, the first evidence of neutrino interactions at LHC was announced and the event  display is shown in Fig.~\ref{fig:FASER}.
	
	\section{Conclusions}
	
	Neutrino had, have and will have a lot to say about the strong dynamics.
	We have recalled the importance of neutrino data in giving physics reality to the fractional
	charge of quarks and recently, the IceCube observations informed about strong dynamics with the neutrino
	telescope in the Antarctic.
	By comparing the rate for up-going and
	down-going neutrino events detected by IceCube, one can  disentangle effects from the unknown
	flux and those from strong (QCD) dynamics.
	Current experimental
	results from IceCube provide  constraints on the
	flux cross-section parameter space. The IceCube HESE data provide
	a measurement of the neutrino-nucleon cross section at
	$\sqrt{s} \sim 1~{\rm TeV}$ that  is consistent within $1\sigma$ with
	perturbative QCD calculations constrained by HERA measurements.
	The data have also
	constrained contributions from non-perturbative processes to the
	neutrino-nucleon cross section. In fact, the contributions to
	the NC interaction at $\sqrt{s} \sim 1~{\rm TeV}$ from electroweak sphaleron
	transitions are excluded at the $2\sigma$ level.
	
	The potential of future neutrino-detection facilities as IceCube-Gen2 for measuring
	the neutrino-nucleon cross section
	shows a clear  improvements to determine both astrophysical neutrino fluxes and cross
	section that can arrive to a precision of about a $6\%$,  comparable to perturbative QCD informed by HERA data.
	Moreover,  a $100~{\rm km}^3$ detector would reach the
	precision of less than a 2\% level.

	\section*{Acknowledgements}
	
	I would like to warmly acknowledge Alberto Santoro for his kind invitation to participate in LISHEP 2021.
	I also thanks to all the organizers of this enjoyable meeting. My gratitude to my colleagues M.B. Gay, J.A. Martins Simoes,
	L. Anchordoqui and J. Soriano for their generous attitude in sharing with me their scientific findings and particularly to L. Anchordoqui
	for helping me in improving the manuscript.

\end{document}